\documentclass[pra,twocolumn,preprintnumbers,amsmath,amssymb]{revtex4}
\usepackage{graphicx}
\usepackage{latexsym}
\usepackage{amsmath}
\usepackage{graphics}
\usepackage{amssymb}
\usepackage{layout}
\usepackage{verbatim}
\usepackage{amsfonts,epsfig}
\newcommand{\ket}[1]{\left|#1\right>}

\newcommand{\beq}{\begin{equation}}
\newcommand{\eeq}{\end{equation}}
\newcommand{\bea}{\begin{eqnarray}}
\newcommand{\eea}{\end{eqnarray}}
\newcommand{\nn}{\nonumber}

\begin{document}

\title{Exactly solvable two-level quantum systems and Landau-Majorana-St\"uckelberg-Zener interferometry}
\author{Edwin Barnes}
\affiliation{Condensed Matter Theory Center, Department of Physics, University of Maryland, College Park, MD 20742-4111, USA}

\begin{abstract}
A simple algorithm is presented based on a type of partial reverse-engineering that generates an unlimited number of exact analytical solutions to the Schr\"odinger equation for a general time-dependent two-level Hamiltonian. I demonstrate this method by deriving new exact solutions corresponding to fast control pulses that contain arbitrarily many tunable parameters. It is shown that the formalism is naturally suited to generating analytical control protocols that perform precise non-adiabatic rapid passage and Landau-Zener interferometry near the quantum speed limit. A general, exact formula for Landau-Zener interference patterns is derived.
\end{abstract}

\maketitle

\section{Introduction}

Although they have pervaded quantum physics since its inception, very few time-dependent two-level quantum systems are known to be analytically solvable. Among the most famous examples of exactly soluble two-level evolution is the Landau-Majorana-St\"uckelberg-Zener (LMSZ) problem \cite{Landau_PZS32,Zener_PRSL32,Stuckelberg_HPA32,Majorana_NC32,Shevchenko_PR10}, which remains a very active area of research due to numerous applications pertaining to quantum phase transitions \cite{Zurek_PRL05}, quantum control \cite{Rudner_PRL08,Petta_Science10,Ribeiro_arxiv1210,Quintana_PRL13,Zhang_arxiv12} and quantum state preparation \cite{Wu_PRL11,Brierley_PRL12,Malossi_PRA13}. The hyperbolic secant pulse of Rosen and Zener \cite{Rosen_PR32} has played an important role in self-induced transparency \cite{McCall_PR69} and qubit control \cite{Economou_PRB06,Greilich_NP09,Poem_PRL11}, and it has since been found to belong to a larger family of analytical controls \cite{Bambini_PRA81,Bambini_PRA84,Hioe_PRA84,Zakrzewski_PRA85,Silver_PRA85,Robinson_PRA85,Ishkhanyan_JPhysA00,Carmel_PRA00,Kyoseva_PRA05,Vitanov_NJP07,Hioe_Wiley07}. Several of these examples have proven very beneficial to the fields of quantum control and computation \cite{Economou_PRB06,Greilich_NP09,Economou_PRB12,Motzoi_PRL09,Chow_PRA10,Gambetta_PRA11}, where analytical solutions are often central in the design of control fields that are fast, precise, and robust against noise. However, the rarity of such solutions has severely limited one's options in developing an analytical approach to qubit gate design.

In a recent work \cite{Barnes_PRL12}, a systematic method for deriving arbitrarily many families of exactly solvable two-state systems was presented, vastly extending the number of known analytical solutions. This method allows one to input many of the basic features of the desired control field and then compute exactly the corresponding evolution of the system with the provided formulas. However, a limitation of this work is that it applies only to systems where the driving is along a single axis of the Bloch sphere, such as in the case of electrically-driven singlet-triplet qubits \cite{Petta_Science05,Foletti_NP09,Maune_Nature12,Wang_NatComm12}, making it inapplicable to the majority of driven two-level systems.

In this paper, I address this limitation by presenting a method to generate arbitrarily many families of solutions in the most general case where the two-level Hamiltonian has time-dependence along any set of axes of the Bloch sphere. Of course, one can easily generate exactly solvable Hamiltonians by first choosing the evolution operator and then differentiating to obtain the corresponding Hamiltonian, but it is challenging to arrive at a physically meaningful Hamiltonian in this way. In contrast, the method presented here allows one to specify the basic form and many features of the Hamiltonian whose evolution one wishes to solve before proceeding to compute the exact solution for this evolution. This method has important applications in a vast range of problems, including the development of quantum controls for essentially any quantum computing platform and control protocols for performing LMSZ interferometry and non-adiabatic rapid passage (NARP). I illustrate this by deriving new, exactly solvable LMSZ driving fields and control pulses that execute a desired evolution at speeds approaching the quantum speed limit (QSL) \cite{Mandelstam_JPhys45,Bhattacharyya_JPA83,Margolus_PD98,Giovannetti_PRA03,Caneva_PRL09,Bason_NP12,Hegerfeldt_arxiv1305}. Attaining fast evolution times is especially crucial in quantum computing where quantum gates need to be performed on timescales much shorter than the decoherence time. In the case of periodic driving through a level anti-crossing, I show that the formalism allows one to easily derive analytical expressions for LMSZ interference patterns and conditions for coherent destruction of tunneling \cite{Grossmann_PRL91,Stehlik_PRB12}.

\section{Analytically solvable Hamiltonians}

The Hamiltonian we consider has the general form
\beq
H=b_x(t)\sigma_x+b_y(t)\sigma_y+b_z(t)\sigma_z,\label{ham}
\eeq
where the $b_k(t)$ are real functions and the $\sigma_k$ are Pauli matrices. This Hamiltonian describes any time-dependent two-level system, with the functions $b_k(t)$ interpreted as either driving fields or time-dependent energy splittings. Alternatively, we may parametrize the Hamiltonian in terms of rotating-frame fields $\alpha$, $\beta$, and $\varphi$, where $\beta{}e^{i\varphi}{\equiv}{}b_x{+}ib_y$ and $\alpha(t){\equiv}{}2\int_0^tdt'b_z(t'){-}\varphi(t)$. In the appendix, it is shown that one can systematically find analytical solutions for the evolution operator generated by Hamiltonian (\ref{ham}) with $\varphi$ and either $\alpha$ or $\beta$ chosen as desired; although one cannot choose both $\alpha$ and $\beta$ at will (were this the case, all two-state problems could be solved analytically), one still has a large amount of control over the features of the second, unspecified function. For concreteness, we suppose that one wishes to fix $\beta(t)$ at the outset (the formalism can easily be modified to fix $\alpha(t)$ instead). While we cannot then find analytical solutions for arbitrary $\alpha$, there exists a different parametrization of the Hamiltonian in which $\alpha$ is replaced with a new function, $\chi(t)$, such that one can systematically generate an analytical expression for the evolution operator for arbitrary choices of $\beta$, $\varphi$ and $\chi$. Parametrizing the lab-frame evolution operator as
\beq
U=\left(\begin{matrix}u_{11} & -u_{21}^*\cr u_{21} & u_{11}^*\end{matrix}\right), \qquad |u_{11}|^2+|u_{21}|^2=1,\label{defofU}
\eeq
the explicit $u_{11}$, $u_{21}$ and driving fields are (see the appendix)
\bea
u_{11}&=&\cos\chi e^{i\xi_--i\varphi/2},\quad u_{21}=i\eta\sin\chi e^{i\xi_++i\varphi/2},\label{evolfromchi}\\
\xi_\pm&=&\int_0^tdt'\beta\sqrt{1-{\dot\chi^2\over\beta^2}}\csc(2\chi)\pm{1\over2}\sin^{-1}\left({\dot\chi\over\beta}\right)\pm\eta{\pi\over4},\nn\\
b_x&=&\beta\cos\varphi,\quad b_y=\beta\sin\varphi,\nn\\
b_z&{=}&{\ddot\chi{-}\dot\chi\dot\beta/\beta\over2\beta\sqrt{1{-}\dot\chi^2/\beta^2}}{-}\beta\sqrt{1{-}\dot\chi^2/\beta^2}\cot(2\chi){+}\frac{\dot\varphi}{2}.\label{bfromchi}
\eea
The initial conditions $u_{11}(0){=}1$, $u_{21}(0){=}0$ imply $\chi(0){=}0$, and $\dot\chi(0){=}{-}\eta\beta(0)$ ensures that $b_z(0)$ is finite, where $\eta{=}{\pm}1$. Eqs.~(\ref{evolfromchi}),(\ref{bfromchi}) embody one of the main results of this paper, as they constitute a general analytic solution of the evolution generated by the Hamiltonian of Eq.~(\ref{ham}). The task of finding analytical solutions has been reduced to first choosing $b_x$, $b_y$ by picking $\beta$ and $\varphi$ at will. One then selects $\chi$ to produce a desired $b_z$ via Eq.~(\ref{bfromchi}), fixing the Hamiltonian. Once these choices are made, an analytical expression for the evolution operator follows from Eq.~(\ref{evolfromchi}). A simple case is $\chi{=}{-}\eta\int_0^tdt'\beta(t')$ and $\varphi{=}0$, where Eq.~(\ref{bfromchi}) gives that $b_y{=}b_z{=}0$, corresponding to an $x$-rotation for any $\beta$. Another simple example arises when $\beta{=}\chi{=}0$, for which Eq.~(\ref{evolfromchi}) yields a $z$-rotation for any $b_z$.

\section{Quantum speed limit}

In Eqs.~(\ref{evolfromchi}),(\ref{bfromchi}), it is clear that proper solutions necessarily satisfy $|\dot\chi|{\le}|\beta|$. The physical origin of this constraint lies in the notion of the quantum speed limit \cite{Mandelstam_JPhys45,Bhattacharyya_JPA83,Margolus_PD98,Giovannetti_PRA03,Caneva_PRL09,Bason_NP12,Hegerfeldt_arxiv1305}, which refers to the minimum time it takes a quantum state to evolve to a different state in the Hilbert space due to energy-time uncertainty. Indeed, $|\dot\chi|{\le}|\beta|$ implies that the fastest possible evolution from $\chi(0){=}0$ to a desired final value $\chi_{target}{\equiv}\chi(T){>}0$ is obtained by choosing $\chi(t){=}\int_0^tdt'|\beta(t')|$, with the shortest time given by substituting $t{=}T$ in this expression and solving for $T$ in terms of $\chi_{target}$ and whatever parameters might appear in $\beta$. For constant $\beta{=}\beta_0{>}0$, we immediately obtain $T_{min}{=}\chi_{target}/\beta_0$, which is the QSL time $T_{QSL}$ for states evolving under an arbitrary time-independent Hamiltonian in the ``Heisenberg regime" \cite{Giovannetti_PRA03}. We refer to $|\dot\chi|{\le}|\beta|$ as the QSL constraint. The present work leads to a general definition of $T_{QSL}$, $\chi(T_{QSL}){=}\chi_{target}{\equiv}\int_0^{T_{QSL}}dt'|\beta(t')|$, for arbitrary time-dependent two-level systems. This definition is consistent with that used in Ref.~\cite{Caneva_PRL09} for a certain class of time-dependent Hamiltonians. Notice that the QSL evolution $\chi{=}{\pm}\int_0^tdt'\beta(t')$ coincides with $b_z{=}\dot\varphi/2$, suggesting that the fastest quantum operations are those which tend to minimize $b_z{-}\dot\varphi/2$, a tendency that is borne out in the examples given below.

The fact that the QSL appears as a simple condition on $\chi$ makes the formalism of Eqs.~(\ref{evolfromchi}),(\ref{bfromchi}) very effective for designing quantum controls that operate near the QSL. To see how this works for a general $\beta(t)$, note that a simple way to construct a function $\chi(t)$ which obeys the QSL constraint is to first find a function which satisfies the constraint in the case where $\beta(t){=}\beta_0$ is a constant. Denoting this latter function by $\chi_0(t)$ and defining $B(t){\equiv}\int_0^tdt'\beta(t')$, if we choose $\chi(t){=}\chi_0(B(t)/\beta_0)$, then $|\dot\chi|{=}|\beta\chi_0'(B/\beta_0)/\beta_0|{\le}|\beta|$ automatically follows. Note that all the single-axis driving examples of \cite{Barnes_PRL12}, where the notation there is related to the present by $q{=}\cos(2\chi)$, can be extended to multi-axis solutions using this trick. Furthermore, if the control field corresponding to $\chi_0$ operates near the QSL, this will also tend to be the case for the one generated by $\chi$. Focusing then on the case $\beta(t){=}\beta_0$, we can construct controls that operate near the QSL by choosing a $\chi(t)$ which contains parameters that can be tuned to values where $\chi{=}{\pm}\beta_0t$. An important feature of solutions generated from a $\chi(B)$ whose time dependence arises only through $B$ is that the evolution operator is an ordinary function of $B$ and $\varphi$, namely
\beq
\xi_\pm(B){=}{\int_0^B}dB'\sqrt{1{-}\chi'^2}\csc(2\chi){\pm}\frac{1}{2}\sin^{{-}1}(\chi'){\pm}\eta\frac{\pi}{4},\label{xiofB}
\eeq
where $\chi'(B){=}d\chi/dB$. This fact greatly facilitates the design of a desired evolution since one can directly control the values of $u_{11}$ and $u_{21}$ by adjusting $B$ and $\varphi$.

\section{Pulse examples}

\subsection{Gaussian-like pulses}

\begin{figure}
\begin{center}
\includegraphics[height=3.5cm, width=1\columnwidth]{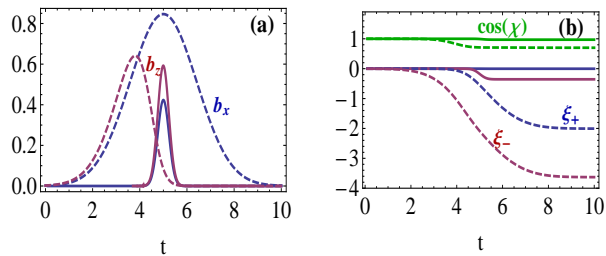}
\caption{\label{fig:pulseexample0} (a) Driving fields from Eq.~(\ref{example0explicit}) and (b) corresponding evolution operator parameters for $t_0{=}5$ and $\mu{=}1/4,\nu{=}3$ (solid), $\mu{=}3,\nu{=}1/2$ (dashed).}
\end{center}
\end{figure}

To illustrate this method of obtaining multi-axis solutions from single-axis ones, consider the example $\chi{=}{-}\frac{1}{2}\cos ^{-1}(e^{-2\beta_0^2t^2})$ where $\beta{=}\beta_0$ and $\eta{=}1$, and the QSL constraint is satisfied. Using the method described above, we can extend this to a solution for any $\beta(t)$:
\beq
\chi=-\frac{1}{2} \cos ^{-1}\left(e^{-2 B(t)^2}\right),\label{example0}
\eeq
which yields the following driving terms:
\bea
&&b_x=\dot B\cos\varphi,\quad b_y=\dot B\sin\varphi,\nn\\
&&b_z=\frac{4 B^2 \dot B}{\sqrt{e^{4 B^2}-4 B^2-1}}+\dot\varphi/2.
\eea
The evolution for any $B$ and $\varphi$ is given by Eqs.~(\ref{evolfromchi}) and (\ref{xiofB}). An explicit example is $B{=}\mu\text{erf}(\nu t)/2$ and $\varphi{=}0$, yielding a Gaussian and a quasi-Gaussian pulse for $b_x,b_z$:
\beq
b_x{=}\frac{\mu\nu e^{-\nu^2\tau^2}}{\sqrt{\pi}},\; b_z{=}\frac{\mu^3\nu e^{{-}\nu^2\tau^2} \text{erf}(\nu \tau)^2/\sqrt{\pi}}{
   \sqrt{e^{\mu^2\text{erf}(\nu\tau)^2}{-}\mu^2\text{erf}(\nu\tau)^2{-}1}},\label{example0explicit}
\eeq
where $\tau{\equiv}t{-}t_0$ and the heights, widths and centers of these pulses can be controlled by tuning the parameters $\mu$ and $\nu$. These pulses and the evolution they produce are shown in Fig.~\ref{fig:pulseexample0}, where it is clear that these pulses have a simple, smooth shape.

\subsection{Pulses with arbitrarily many parameters}

\begin{figure}
\begin{center}
\includegraphics[height=3.7cm, width=1\columnwidth]{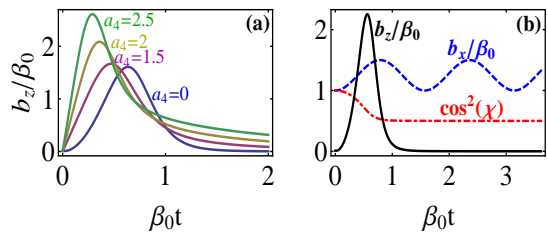}
\caption{\label{fig:pulseexample1} Control field $b_z$ generated by the $\chi$ from (a) Eq.~(\ref{example1a}) with $b_x{=}\beta_0$, $\varphi{=}0$, $k{=}6$, $a_2{=}0$, $a_6{=}4/(\pi\beta_0)$, (b) Eq.~(\ref{example1b}) with $b_x{=}\beta_0(1+\sin^2(2\beta_0t)/2)$, $\varphi{=}0$, $k{=}6$, $a_2{=}a_4{=}0$, $a_6{=}4/\pi$. A Hadamard gate is achieved for total evolution duration $T{=}3.61/\beta_0$.}
\end{center}
\end{figure}

For a near-QSL pulse example, consider the case
\beq
\chi=-\beta_0t[1+(a_2t)^2+(a_4t)^4+\dots+(a_kt)^k]^{-1/k},\label{example1a}
\eeq
where $\beta(t){=}\beta_0$ and the $a_i$ are arbitrary constants, $k$ is an even integer, and $\varphi{=}0$, $\eta{=}1$. The QSL constraint is satisfied regardless of how large $k$ is, so that this $\chi$ yields an exact solution with arbitrarily many parameters $a_i$. We can make the corresponding control field a pulse by setting $a_k{=}4\beta_0/\pi$, so that $\chi{\to}{-}\pi/4$ and $b_z{\to}0$ as $t{\to}\infty$. The initial value of the pulse is set by $a_2$: $b_z(0){=}2a_2\beta_0$. Examples of these pulses are shown in Fig.~\ref{fig:pulseexample1}a. The duration of the pulse approaches $T_{QSL}$ in the limit $a_{i<k}{\to}0$, $k{\to}\infty$, as can be seen by observing that $\chi{\to}{-}\beta_0t$ in this limit. The substantial amount of tunability in this solution already makes it very attractive for applications in quantum computation such as dynamically corrected gates \cite{Motzoi_PRL09,Gambetta_PRA11,Wang_NatComm12}, where the shape of the pulse is tuned to perform a specific quantum operation while simultaneously suppressing errors.

Using the prescription outlined above, we can extend this solution to the case of non-constant $\beta$:
\beq
\chi=-B(t)[1+(a_2B(t))^2+(a_4B(t))^4+\dots+(a_kB(t))^k]^{-1/k}.\label{example1b}
\eeq
This class of pulses can be used to implement quantum operations by tuning $b_z(t)$ for a given choice of $b_x(t)$ and $b_y(t)$. We demonstrate this by designing a fast $b_z$ pulse that, together with $b_x$, implements a Hadamard gate, a quantum operation that is ubiquitous in the field of quantum information processing and which is equivalent to a $\pi$-rotation about $\hat x{+}\hat z$. First choose $a_k{=}4/\pi$, which ensures that $|u_{11}|,|u_{21}|{\to}1/\sqrt{2}$. Supposing $\varphi{=}0$, if we let the system evolve for a time $T$ such that $\int_0^{B(T)} dB'\sqrt{1{-}\chi'^2}\csc(2\chi){=}{-}5\pi/4$, then the phases of $u_{11}$ and $u_{21}$ will also attain their Hadamard values. Such a $b_z$ pulse is shown in Fig.~\ref{fig:pulseexample1}b for an oscillatory $b_x$. From Fig.~\ref{fig:pulseexample1}b, it is evident that $b_z$ quickly sets the magnitudes of $u_{11}$ and $u_{21}$, while the remainder of the evolution with $b_z{\approx}0$ allows their phases to accumulate. As before, the duration of the pulse approaches $T_{QSL}$ as $a_{i<k}{\to}0$, $k{\to}\infty$. This example illustrates how one can use this formalism to design analytical controls near the QSL in the presence of additional driving fields.

\section{LMSZ interferometry}

\begin{figure}
\begin{center}
\includegraphics[height=3cm, width=0.5\columnwidth]{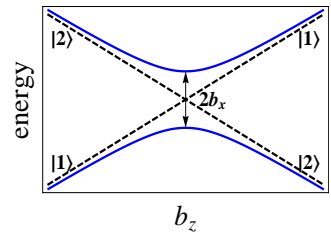}
\caption{\label{fig:LZenergies} Diabatic and adiabatic energy levels.}
\end{center}
\end{figure}
The present formalism is also natural for designing driving fields that perform controlled LMSZ interferometry and non-adiabatic rapid passage, phenomena which have many applications in quantum control \cite{Petta_Science10,Ribeiro_arxiv1210,Quintana_PRL13,Zhang_arxiv12}, state preparation \cite{Wu_PRL11,Brierley_PRL12} and qubit readout \cite{Petta_Science05,Foletti_NP09,Maune_Nature12}. (See Refs.~ \cite{Berry_PRSL90,Lim_JPA91,Vitanov_PRA99,Berry_JPA09,Malossi_PRA13,Ruschhaupt_arxiv12} and references therein for previous analytical approaches for the LMSZ problem.) The LMSZ problem is generally setup as follows. Define the eigenstates of $\sigma_z$ to be $\ket{1}$ and $\ket{2}$ and set $\varphi{=}0$ so that $\beta{=}b_x$; when $|b_z|{\gg}|b_x|$, these states are approximate energy eigenstates. A nonzero $b_x$ produces an anti-crossing with an energy gap of $2b_x$ (see Fig.~\ref{fig:LZenergies}) which may be time-dependent. Now suppose that we drive $b_z$ through the anti-crossing, starting from some large negative value at $t{=}0$ up to a large positive value at $t{=}T$. Assuming that the system is initially prepared in state $\ket{1}$ at time $t{=}0$, the probability $P_2(T)$ for the system to be in state $\ket{2}$ at time $t{=}T$ is
\beq
P_2(T)=|u_{21}(T)|^2=\sin^2[\chi(T)].\label{NARPprob}
\eeq
The fact that this depends only on $\chi(T)$ demonstrates the suitability of the present formalism to the LMSZ problem. If we choose $\chi$ such that $\chi(0){=}0$ and $\chi(T){=}0$, then we achieve a perfect LMSZ transition: the system is driven through the anti-crossing and returns to state $\ket{1}$ with probability 1. On the other hand, we may choose $\chi(T){=}\pi/2$, in which case the system undergoes NARP and ends up in state $\ket{2}$ after being driven through the anti-crossing. Other values of $\chi(T)$ lead to superpositions of $\ket{1}$ and $\ket{2}$. We may also consider LMSZ interferometry, where the system is driven through the anti-crossing periodically, and the resulting time-averaged probabilities to be in state $\ket{1}$ or $\ket{2}$ after many periods is again largely determined by $\chi(T)$, as we will see. In choosing a $\chi(t)$ for the LMSZ problem, we must impose appropriate initial conditions. For simplicity, we focus on the case $b_z(0){=}{-}\infty$, $b_z(T){=}\infty$, for which we need $\ddot\chi(0)<0$, $\ddot\chi(T)>0$; the analysis can be extended straightforwardly to the case where $b_z$ is finite at $t{=}0,T$.

\begin{figure}
\begin{center}
\includegraphics[height=6cm, width=\columnwidth]{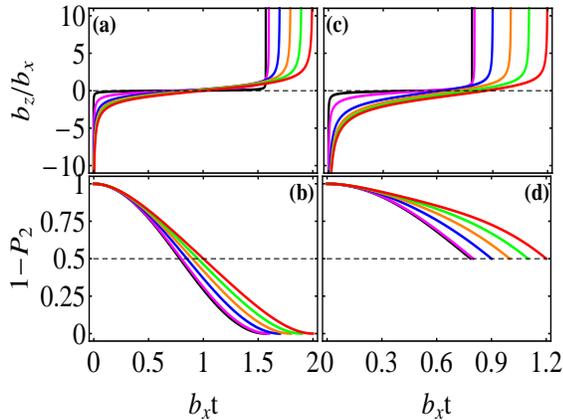}
\caption{\label{fig:example1} (a) Control field and (b) NARP probability from Eqs.~(\ref{example1chi}),(\ref{NARPprob}),(\ref{bfromchi}) with $\chi(T){=}\pi/2$ and $b_xT{=}\pi/2$,1.6,1.7,1.8,1.9,2 and (c),(d) $\chi(T){=}\pi/4$ and $b_xT{=}\pi/4,0.8,0.9,1,1.1,1.2$.}
\end{center}
\end{figure}
For constant $b_x$, a simple example which satisfies these boundary conditions and the QSL constraint is
\beq
\chi=b_xt-{ab_xT\over2}t^2+{ab_x\over3}t^3,\label{example1chi}
\eeq
where choosing $0{\le} a{\le} 16/(3T^2)$ ensures that the $b_z$ from Eq.~(\ref{evolfromchi}) is finite in the interval $t{\in}(0,T)$, and $\eta{=}{-}1$. This $\chi$ saturates the QSL constraint when $a{=}0$, implying that $b_z$ will implement near-QSL evolution for small $a$. To achieve a target $\chi(T)$, set $t{=}T$ in Eq.~(\ref{example1chi}) and solve for $a$: $a(T){=}6[b_xT{-}\chi(T)]/(b_xT^3)$. Plugging this into Eqs.~(\ref{example1chi}) and (\ref{bfromchi}) yields a family of driving fields $b_z$ parametrized by $T$ that achieve the desired evolution for any $\chi(T){\in}(0,\pi/2]$; some of these are shown in Fig.~\ref{fig:example1} along with the corresponding NARP probabilities. The restrictions on $a(T)$ impose bounds on $T$: $\chi(T){\le} b_xT{\le} 9\chi(T)$; the upper bound is particular to Eq.~(\ref{example1chi}), while the lower bound is the familiar, universal QSL and gives rise to the step-like curves in Figs.~\ref{fig:example1}a and \ref{fig:example1}c. These curves reveal that the desired LMSZ transition is achieved as quickly as possible by first driving $b_z$ to zero very rapidly, allowing the system to evolve for a time $T\lesssim T_{QSL}$, and then driving $b_z$ quickly up to its final value (see also Ref.~\cite{Hegerfeldt_arxiv1305}). In addition to NARP, these near-QSL driving fields could be important for LMSZ-based generation of entanglement in superconducting qubits \cite{Quintana_PRL13}, where fidelities are often limited by short relaxation times.

\begin{figure}
\begin{center}
\includegraphics[height=3.5cm, width=0.7\columnwidth]{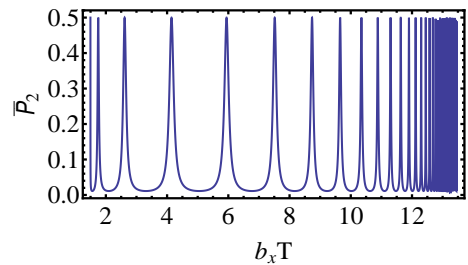}
\caption{\label{fig:example2} Time-averaged excited state probability $\bar{\cal P}_2$ from Eq.~(\ref{avgprob}) with periodic driving field generated by Eq.~(\ref{example1chi}) with $\chi(T){=}\pi/2.1$ and half-period $T$. The non-monotonicity in the fringe spacing reflects the non-monotonicity of Eq.~(\ref{example1chi}) at larger values of $T$.}
\end{center}
\end{figure}
In the context of LMSZ interferometry, the formalism of Eqs.~(\ref{evolfromchi}),(\ref{bfromchi}) yields an exact formula for LMSZ interference patterns. To show this, we begin by constructing a periodic driving field $b_z$ of period $2T$ where $\chi(t)$ determines the first half of one period and $\chi_2(t){=}\chi(2T{-}t)$ the second half, corresponding to $b_z$ retracing its path. Using Eq.~(\ref{evolfromchi}), we find the evolution after one full period:
\bea
u_{11}(2T)&=&e^{2i\xi_0(T)}\cos(2\chi(T)),\nn\\ u_{21}(2T)&=&-i\sin(2\chi(T)),
\eea
where $\xi_0{\equiv} (\xi_{+}{+}\xi_{-})/2$ and we have assumed that $\dot\chi(T){=}{-}\eta\beta(T)$ for simplicity. From this expression, it is straightforward to compute the time-averaged probability of being in the excited state $\ket{2}$ after many periods:
\beq
\bar{\cal P}_2=[2+2\cot^2(2\chi(T))\sin^2(2\xi_0(T))]^{-1}.\label{avgprob}
\eeq
Thus, we see that the present formalism readily produces a general, exact, analytic formula for $\bar{\cal P}_2$, whereas analytic expressions for this important quantity typically require several approximations \cite{Shevchenko_PR10}. This function takes values in the range $[0,1/2]$, where $\bar{\cal P}_2{=}1/2$ for $\chi(T){=}\pi/4$, while $\bar{\cal P}_2{=}0$ for $\chi(T){=}\pi/2$. This is to be expected since $\chi(T){=}\pi/4$ corresponds to a 50-50 ``beam splitter", while $\chi(T){=}\pi/2$ ensures the system remains in the ground state after every sweep through the anti-crossing. In the context of charge qubits where states $\ket{1}$ and $\ket{2}$ correspond to an electron being in either the left or right dot of a tunnel-coupled double quantum dot, the case $\bar{\cal P}_2{=}0$ can be interpreted as coherent destruction of tunneling \cite{Grossmann_PRL91,Stehlik_PRB12} since the electron becomes localized in one dot. For more generic values of $\chi(T)$, $\bar{\cal P}_2$ is modulated by the phase $\xi_0(T)$, which can produce interference fringes as control parameters are adjusted, as shown in Fig.~\ref{fig:example2} for the example from Eq.~(\ref{example1chi}) with $\chi(T){=}\pi/2.1$. As $T$ is varied, an interference pattern emerges in which the peaks of the pattern sharpen and eventually disappear as $\chi(T)$ approaches $\pi/2$. Interestingly, Fig.~\ref{fig:example2} reveals a peak at the QSL time $T{=}T_{QSL}{=}\pi/(2.1b_x)$; this is generally the case since at the QSL, $|\dot\chi|{=}|\beta|$, so that $\xi_0(T_{QSL}){=}0$. This leads to the surprising conclusion that if we choose $\chi(T){=}\pi/2$ in order to trap the system in state $\ket{2}$, then driving very close to the QSL may not be ideal since small deviations away from $\chi(T){=}\pi/2$ will produce the peak at $T{=}T_{QSL}$ and, hence, large uncertainty in the state of the system.

\section{Conclusions}

In conclusion, a general formalism for deriving exactly solvable time-dependent two-level quantum systems has been presented. This formalism can vastly increase the number of known exact solutions for physical Hamiltonians, as has been demonstrated with explicit examples. These examples show that this theory can be a powerful tool in the design of control pulses both for quantum computation and for precise Landau-Majorana-St\"uckelberg-Zener interferometry near the quantum speed limit.

\bigskip
\centerline{\bf Acknowledgments}
\bigskip

I thank Michael Berry, Lev Bishop, {\L}ukasz Cywi\'nski, and Sophia Economou for helpful discussions. I also thank Gerhard Hegerfeldt for useful comments. This work is supported by LPS-NSA and IARPA.

\appendix

\section{}

In this appendix, we derive Eqs.~(\ref{evolfromchi}) and (\ref{bfromchi}) of the main text. The general form of the Hamiltonian
and its corresponding evolution operator are given in Eqs.~(\ref{ham}),(\ref{defofU}). The evolution operator obeys a Schr\"odinger equation whose form can be made compact by transforming to a rotating frame: $v_{11}=e^{i\int_0^tdt'b_z(t')}u_{11}$, $v_{21}= e^{-i\int_0^tdt'b_z(t')}u_{21}$. Defining $\beta e^{i\varphi}\equiv b_x+ib_y$ and $\alpha(t)\equiv 2\int_0^tdt'b_z(t')-\varphi(t)$, we have
\beq
\dot v_{11}=-i\beta e^{i\alpha}v_{21},\qquad \dot v_{21}=-i\beta e^{-i\alpha}v_{11}.\label{vschrod}
\eeq
Here, it is manifest that the evolution operator in the rotating frame depends on only two real functions, $\alpha$ and $\beta$. We must further specify $b_z$ to return to the lab frame, but this choice can be made after the evolution operator is computed in the rotating frame. In what follows, I will show that one can systematically find analytical solutions with either $\alpha$ or $\beta$ chosen as desired; although one cannot choose both $\alpha$ and $\beta$ at will (if this were the case, all two-state problems would be analytically solvable), we will see that one still has a large amount of control over the features of the second, unspecified function.

For concreteness, we will suppose that $\beta(t)$ is chosen at the outset (the formalism can easily be modified to fix $\alpha(t)$ instead). While we cannot then find analytical solutions for arbitrary $\alpha$, there exists a different parametrization of the Hamiltonian in which $\alpha$ is replaced with a new function, $\kappa_I(t)$, such that one can systematically generate an analytical expression for the evolution operator for arbitrary choices of $\beta$ and $\kappa_I$. To see this, first express the rotating-frame evolution operator in terms of some complex $\kappa(t)$:
\beq
v_{11}=e^{-i\int_0^tdt'\beta(t')e^{\kappa(t')}},\quad v_{21}=-i\eta e^{-i\int_0^tdt'\beta(t')e^{-\kappa(t')}},\label{vfromkappa}
\eeq
with $\eta=\pm1$. This choice of parametrization is motivated by observing that we can combine the two equations in (\ref{vschrod}) to obtain $\beta^2=-(\dot v_{11}/ v_{11})(\dot v_{21}/ v_{21})$, which generally implies $\dot v_{11}/ v_{11}=-i\beta e^\kappa$, $\dot v_{21}/ v_{21}=-i\beta e^{-\kappa}$ for some complex $\kappa$. The fact that this is true in general can be seen by noting that for $\beta\ne0$, any complex function can be expressed as $-i\beta e^{\kappa}$ for some complex function $\kappa$, so that we may therefore express $\dot v_{11}/ v_{11}$ in this way. $\beta^2=-(\dot v_{11}/ v_{11})(\dot v_{21}/ v_{21})$ then implies $\dot v_{21}/ v_{21}=-i\beta e^{-\kappa}$. This argument does not hold when $\beta=0$, however in this case we find $\dot v_{11}=0=\dot v_{21}$, which is consistent with Eq.~(\ref{vschrod}). This analysis is thus completely general and applies to any solution of the Schr\"{o}dinger equation. Consistency between Eqs.~(\ref{vfromkappa}) and (\ref{vschrod}) requires
\beq
\alpha(t)=-i\kappa(t)+\eta{\pi\over2}-2\int_0^tdt'\beta(t')\sinh\kappa(t'),\label{alphafromkappa}
\eeq
which should be interpreted as follows: For any choice of a complex $\kappa(t)$ and real $\beta(t)$ such that the $\alpha(t)$ computed from Eq.~(\ref{alphafromkappa}) is real, the evolution operator obtained from Eq.~(\ref{vfromkappa}) is the exact solution for this $\alpha$ and $\beta$. Writing $\kappa=\kappa_R+i\kappa_I$ and imposing $\hbox{Im}(\alpha)=0$ determines $\kappa_R$ in terms of $\kappa_I$: $\kappa_R=-2\tanh^{-1}\tan\left(\chi+\pi/4\right)$, with
\beq
\chi(t)\equiv\int_0^tdt'\beta(t')\sin(\kappa_I(t')),\label{defofchi}
\eeq
leading to an expression for $\alpha$ that is real for any $\kappa_I$:
\beq
\alpha=\kappa_I+\eta{\pi\over2}-2\int_0^tdt'\beta(t')\cos\kappa_I(t')\cot\left[2\chi(t')\right].\label{alphafromkappaI}
\eeq

While this parametrization has the nice feature that $\kappa_I$ can be chosen freely, one drawback is that one must then perform the integration in Eq.~(\ref{defofchi}), making it harder to relate the features of $\kappa_I$ to the driving field $\dot\alpha$. We can avoid this by specifying $\chi$ directly, but at the expense of now having to choose functions $\chi(t)$ that obey the quantum speed limit constraint, $|\dot\chi|\le|\beta|$, which arises directly from Eq.~(\ref{defofchi}). Solving Eq.~(\ref{defofchi}) for $\kappa_I$ in terms of $\chi$, it is straightforward to turn the above expressions for the evolution operator into expressions which depend on $\chi$ rather than $\kappa_I$. The resulting lab-frame evolution operator and driving fields are given in Eqs.~(\ref{evolfromchi}) and (\ref{bfromchi}).

\end{document}